\documentclass []{mn2e}

\title[Eccentricity generation in HTS]{Eccentricity generation in hierarchical triple systems with
coplanar and initially circular orbits}
\author[Nikolaos Georgakarakos]{Nikolaos Georgakarakos\\Department of Mathematics and Statistics, Edinburgh
University\\Mayfield Road, Edinburgh EH9 3JZ\\
email: ng@maths.ed.ac.uk}
\date{}

\usepackage{graphicx}

\begin{document}
\maketitle

\begin{abstract}
We develop a technique for estimating the inner
eccentricity in hierarchical triple systems with well separated
components.  We investigate systems with initially circular and
coplanar orbits and comparable masses.  The technique is based on an
expansion of the rate of change of the Runge-Lenz vector for
calculating short period terms by using first order perturbation
theory.  The combination of the short period terms with terms arising from
octupole level secular theory, results in the derivation of a  rather
simple formula for the eccentricity of the inner binary.  The
theoretical results are tested against numerical integrations of the
full equations of motion. Comparison is also made with other results
on the subject.
\end{abstract}


\section{INTRODUCTION}
A hierarchical triple system consists of a binary system and a third body on a
wider orbit.  The motion of such a system can be pictured as the motion
of two binaries: the binary itself (inner binary) and the binary which
consists of the third body and the centre of mass of the binary (outer binary).  Hierarchical triple systems are widely present in the
galactic field and in star clusters and studying the dynamical evolution of such
systems is a key to understanding a number of issues in astronomy and
astrophysics.  Sometimes, for example, the inner pairs in triple
stellar systems are {\sl close}
binary systems, i.e. the separation between the components is
comparable to the radii of the bodies.  In these circumstances, the
behaviour of the inner binary can depend very sensitively on the
separation of its components and this in turn is affected by the third
body.  Thus, a slight change in the separation of the binary
stars can cause drastic changes in processes such as tidal friction
and dissipation, mass transfer and  mass loss due to a stellar wind,
which may result in changes in stellar
structure and evolution.  Eventually, these physical changes
can affect the dynamics of the whole triple system.  But even in
systems with well-separated inner binary components, the perturbation
of the third body can have a devastating effect on the 
triple system as a whole (e.g. disruption of the system).     

For most hierarchical triple stars, the period ratio
${X}$ is of the order of 100 and these systems are probably very stable
dynamically.  However, there are systems with much smaller period
ratios, like the system HD 109648 with ${X=22}$ (Jha et al. 2000), the
${\lambda}$ Tau system, with
\begin{math}
X=8.3
\end{math}
(Fekel ${\&}$ Tomkin 1982)
and the CH Cyg system with
\begin{math}
X=7.0
\end{math}
(Hinkle et al. 1993).  Our aim is to find how much inner binary eccentricity is
generated in systems with large period ratio ${X}$ (${X>10}$).  We consider the case where the inner eccentricity is initially zero,
since in close binaries tidal friction is expected to circularise the
orbit.  The outer orbit is also circular. 

The initial motivation to the work presented in this paper was given by the
work of Peter Eggleton and his collaborators on stellar and dynamical
evolution of triple systems (Eggleton ${\&}$ Kiseleva 1996, Kiseleva,
Eggleton ${\&}$ Mikkola 1998).  Other recent work on the dynamics of
hierarchical triple system includes the work done by Ford, Kozinsky ${\&}$ Rasio (2000) and Krymolowski ${\&}$ Mazeh (1999).

\section{THEORY}
\label{theo}
We are going to derive expressions for the short period (which varies
on a time-scale comparable to the inner and outer orbital periods) and
secular modulations of the inner eccentricity.  The short period terms
will be obtained in a rather simple way, by using the definition of
the Runge-Lenz vector, while the secular evolution, where it is needed, will be studied by
means of canonical perturbation theory.  It is also possible to obtain
the short period terms by using canonical methods.  However, as seen
in the following section, using the definition of the eccentric vector
is a quite straightforward procedure which does not require any
knowledge of canonical perturbation theory.  

An important aspect of the theory that is developed in the subsequent sections
is the combination of the short period and secular terms in the expressions
for the eccentricities.  At any moment of the evolution of the system,
we will consider that the eccentricity (inner or outer)
consists of a short period and a long period (secular) component,
i.e. ${e=e_{{\rm short}}+e_{{\rm sec}}}$ (one can picture this by
recalling the expansion of the disturbing function in solar system
dynamics, where the perturbing potential is given as a sum of an
infinite number of cosines of various frequencies).  Thus,
considering the eccentricity to be initially zero leads to ${e_{{\rm
short}}=-e_{{\rm sec}}}$ (initially), which implies that, although the
eccentricity is initially zero, the short and secular eccentricity may
not be.

Finally, in this paper, as was stated earlier, we will be
concentrating on systems with well separated components and comparable
masses.  Therefore, while developing the theoretical model in the next
sections, we will consider ${X}$ to be large (or any equivalent form
of that assumption). 

\subsection{Calculation of the short-period contribution to the eccentricity}	
\label{s1}
\begin{figure}
\begin{center}
\includegraphics[width=70mm,height=50mm]{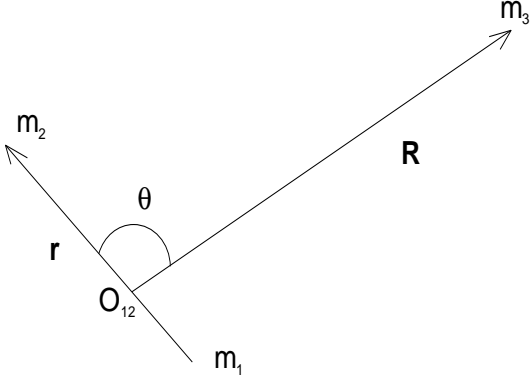}
\caption[]{The Jacobi formulation.  The point ${O_{12}}$ is the centre
of mass of the inner binary.}
\label{jacform}
\end{center}
\end{figure}

First, we calculate the short-period terms.  The motion of the system can be studied using the Jacobi decomposition
of the three-body problem (Fig. \ref{jacform}).  In that context, the equation of motion of
the inner binary is:
\begin{equation}
\ddot{\bmath {r}}=-G(m_{1}+m_{2})\frac{\bmath{r}}{r^{3}}+\bmath{F},
\label{eqmo}
\end{equation}
where ${\bmath{F}}$, the perturbation to the inner binary motion, is  
\begin{eqnarray}
\bmath{F} & = & Gm_{3}(\frac{\bmath{R}-\mu_{1}\bmath{r}}{|\bmath{R}-\mu_{1}\bmath{r}|^{3}}-\frac{\bmath{R}+\mu_{2}\bmath{r}}{|\bmath{R}+\mu_{2}\bmath{r}|^{3}})=\nonumber\\
& = & Gm_{3}\frac{\partial}{\partial{\bmath{r}}}(\frac{1}{\mu_{1}|\bmath{R}-\mu_{1}\bmath{r}|}+\frac{1}{\mu_{2}|\bmath{R}+\mu_{2}\bmath{r}|})
\label{potenc1}
\end{eqnarray}
with
\begin{displaymath}
\mu_{{\rm i}}=\frac{m_{{\rm i}}}{m_{1}+m_{2}}, \hspace{0.5 cm} i=1,2.
\end{displaymath}
Now, since the third star is at considerable distance from
the inner binary, implying that 
\begin{math}
r/R 
\end{math}
is small, the inverse distances in equation (\ref{potenc1}) can be expressed as:
\begin{displaymath}
\frac{1}{|\bmath{R}-\mu_{1}\bmath{r}|}=\frac{1}{R}\sum^{\infty}_{n=o}
\left( \frac{\mu_{1}r}{R} \right) ^{n}P_{{\rm n}}(\cos{\theta})
\end{displaymath}
and
\begin{displaymath}
\frac{1}{|\bmath{R}+\mu_{2}\bmath{r}|}=\frac{1}{R}\sum^{\infty}_{n=o}
\left(- \frac{\mu_{2}r}{R} \right)^{n}P_{{\rm n}}(\cos{\theta}),
\end{displaymath}
where ${P_{{\rm n}}}$ are the Legendre polynomials and ${\theta}$ is the
angle between the vectors ${\bmath{r}}$ and ${\bmath{R}}$.  Expanding to third
order, the perturbation becomes

\begin{displaymath}
\bmath{F}=Gm_{3}\frac{\partial}{\partial{\bmath{r}}}\left(\frac{3}{2}\frac{(\bmath{r}\cdot
\bmath{R})^{2}}{R^{5}}-\frac{1}{2}\frac{r^{2}}{R^{3}}-\frac{5(\mu_{2}^{2}-\mu_{1}^{2})}{2}\frac{(\bmath{r}\cdot\bmath{R})^{3}}{R^{7}}+\right.
\end{displaymath}
\begin{equation}
\left.+\frac{3(\mu_{2}^{2}-\mu_{1}^{2})}{2}\frac{r^{2}(\bmath{r}\cdot\bmath{R})}{R^{5}}\right).
\label{fpert}
\end{equation}
The first two terms in the above equation come from the quadrupole
term (${P_{2}}$), while the other two come from the octupole term
(${P_{3}}$).  

Using now the definition of the eccentric vector,
i.e. the vector which has the same direction as the radius vector to
the pericentre and whose magnitude is equal to the eccentricity of the
orbit, we can obtain an expression for the inner eccentricity.  The
inner eccentric vector
${\bmath{e}_{1}}$ is given by
\begin{equation}
\bmath{e}_{1}=-\frac{\bmath{r}}{r}+\frac{1}{\mu}(\dot{\bmath{r}}\bmath{\times}\bmath{h}),
\label{ecve}
\end{equation}
where
\begin{math}
\bmath{h}=\bmath{r}\bmath{\times}\dot{\bmath{r}}
\end{math}
and
\begin{math}
\mu=G(m_{1}+m_{2}).
\end{math} 
Differentiating equation (\ref{ecve}) and substituting for
${\bmath{F}}$ (we neglect the term
${\bmath{r}\cdot\dot{\bmath{r}}}$ because, for the
applications discussed in this paper, is expected to be  small and of ${O(e)}$), we obtain:
\begin{eqnarray}
\dot{\bmath{e}}_{1} & = & \frac{Gm_{3}}{\mu
R^{3}}\left[\left(6\frac{(\bmath{r}\cdot\bmath{R})(\dot{\bmath{r}}\cdot\bmath{R})}{R^{2}}-\right.\right.\nonumber\\
& & -15(\mu_{2}^{2}-\mu_{1}^{2})\frac{(\bmath{r}\cdot\bmath{R})^{2}(\dot{\bmath{r}}\cdot\bmath{R})}{R^{4}}+\nonumber\\
& &
\left.+3(\mu_{2}^{2}-\mu_{1}^{2})\frac{r^{2}(\dot{\bmath{r}}\cdot\bmath{R})}{R^{2}}\right)\bmath{r}+\nonumber\\
& & +\left(r^{2}-3\frac{(\bmath{r}\cdot\bmath{R})^{2}}{R^{2}}+\frac{15}{2}(\mu_{2}^{2}-\mu_{1}^{2})\frac{(\bmath{r}\cdot\bmath{R})^{3}}{R^{4}}-\right.\nonumber\\
& & \left.\left.-\frac{9}{2}(\mu_{2}^{2}-\mu_{1}^{2})\frac{r^{2}(\bmath{r}\cdot\bmath{R})}{R^{2}}\right)\dot{\bmath{r}}\right].\label{roo5}
\end{eqnarray}
Now, the Jacobi vectors can be represented approximately in polar form as
\begin{math}
\bmath{r}=a_{1}(\cos{n_{1}t},\sin{n_{1}t})
\end{math}
and
\begin{math}
\bmath{R}=a_{2}(\cos{(n_{2}t+\phi)},\sin{(n_{2}t+\phi)})
\end{math}
(again, the terms neglected are of ${O(e)}$), where ${a_{1}}$ and ${a_{2}}$ are the semi-major axes of the inner and
outer orbit respectively and ${\phi}$ is the initial relative phase of the two binaries.
After integrating, the components ${x_{1}}$ and ${y_{1}}$ of the
eccentric vector become (expanding in powers of
\begin{math}
\frac{1}{X}
\end{math}
and retaining the two leading terms):
\begin{equation}
x_{1}=\frac{m_{3}}{M}\frac{1}{X^{2}}(P_{{\rm x21}}(t)+X^{\frac{1}{3}}P_{{\rm x31}}(t))+C_{{\rm
x}_{1}}\label{e11}
\end{equation}
\begin{equation}
y_{1}=\frac{m_{3}}{M}\frac{1}{X^{2}}(P_{{\rm y21}}(t)+X^{\frac{1}{3}}P_{{\rm y31}}(t))+C_{{\rm
y}_{1}}\label{e12}
\end{equation}
where
\begin{eqnarray}
P_{{\rm x21}}(t) & = &
-\frac{1}{2}\cos{n_{1}t}+\frac{1}{4}\cos{((3n_{1}-2n_{2})t-2\phi)}+\nonumber\\
& & +\frac{9}{4}\cos{((n_{1}-2n_{2})t-2\phi)}\\
P_{{\rm x31}}(t) & = & \frac{15}{16}m_{*}\cos{(n_{2}t+\phi)}\\
P_{{\rm y21}}(t) & = & -\frac{1}{2}\sin{n_{1}t}+\frac{1}{4}\sin{((3n_{1}-2n_{2})t-2\phi)}-\nonumber\\
& & -\frac{9}{4}\sin{((n_{1}-2n_{2})t-2\phi)}\\
P_{{\rm y31}}(t) & = & \frac{15}{16}m_{*}\sin{(n_{2}t+\phi)}
\end{eqnarray}
\begin{equation}
m_{*}=\frac{m_{2}-m_{1}}{(m_{1}+m_{2})^{\frac{2}{3}}M^{\frac{1}{3}}}.
\end{equation}
${M}$ is the total mass of the system and ${C_{{\rm x}_{1}}}$ and
${C_{{\rm y}_{1}}}$ are
constants of integration.  The semi-major axes and mean motions were
treated as constants in the above calculation.

\subsection{Calculation of the secular contribution to the eccentricity}	
\label{s2}
In order to derive the long-term modulation of the system, we use a
Hamiltonian which is averaged over the inner and outer orbital periods
by means of the Von Zeipel method.  Secular terms cannot be obtained
by the method of section \ref{s1}, because, for an eccentric outer
binary, those terms appear as a linear function of time in the
expansion of the eccentric vector and therefore, they are valid for limited time.

The doubly averaged Hamiltonian
for coplanar orbits is (Marchal 1990, Krymolowski ${\&}$ Mazeh 1999):

\begin{eqnarray}
H & = &-\frac{Gm_{1}m_{2}}{2a_{{\rm
S}}}-\frac{G(m_{1}+m_{2})m_{3}}{2a_{{\rm T}}}+Q_{1}+\nonumber\\
& & +Q_{2}+Q_{3}, \label{hamilto} \\
\mbox{where}\nonumber\\
Q_{1} & = &
-\frac{1}{8}\frac{Gm_{1}m_{2}m_{3}a^{2}_{{\rm
S}}}{(m_{1}+m_{2})a^{3}_{{\rm T}}(1-e^{2}_{{\rm T}})^{\frac{3}{2}}}(2+3e^{2}_{{\rm S}}), \\
Q_{2} & = &
\frac{15Gm_{1}m_{2}m_{3}(m_{1}-m_{2})a^{3}_{{\rm S}}e_{{\rm S}}e_{{\rm
T}}}{64(m_{1}+m_{2})^{2}a^{4}_{{\rm T}}(1-e^{2}_{{\rm T}})^{\frac{5}{2}}}\times\nonumber\\
& & \times\cos{(g_{{\rm S}}-g_{{\rm T}})}(4+3e^{2}_{{\rm S}}),\label{marq}\\
Q_{3} & = &
-\frac{15}{64}\frac{Gm_{1}m_{2}m_{3}^{2}a_{{\rm
S}}^{\frac{7}{2}}e_{{\rm S}}^{2}(1-e_{{\rm
S}}^{2})^{\frac{1}{2}}}{(m_{1}+m_{2})^{\frac{3}{2}}M^{\frac{1}{2}}a_{{\rm
T}}^{\frac{9}{2}}(1-e_{{\rm T}}^{2})^{3}}\times\nonumber\\
& & \times[5(3+2e_{{\rm T}}^{2})+3e^{2}_{{\rm T}}\cos{2(g_{{\rm
S}}-g_{{\rm T}})}].
\end{eqnarray}
The subscripts S and T refer to the inner and outer long period
orbits respectively, while ${g}$ is used to denote longitude of pericentre.  The first term in the Hamiltonian is the Keplerian energy of the inner
binary, the second term is the Keplerian energy of the outer binary,
while the other three terms represent the interaction between the two
binaries.  The ${Q_{1}}$ term comes from the ${P_{2}}$ Legendre
polynomial, the ${Q_{2}}$ term comes from the ${P_{3}}$
Legendre polynomial and the ${Q_{3}}$ term arises from the canonical transformation.
It should be mentioned here that in Marchal, ${Q_{3}}$ includes only the term which is
independent of the arguments of pericentre and the ${P_{3}}$ term in
Krymolowski and Mazeh has the wrong sign.  The same sign error appears in
Ford, Kozinsky and Rasio.
 
By using Hamilton's equations, we can now derive the averaged
equations of motion of the system.  Hence,
\begin{eqnarray}
\frac{{\rm d}x_{{\rm S}}}{{\rm d}\tau} & = &
\frac{5}{16}\alpha\frac{e_{{\rm T}}}{(1-e^{2}_{{\rm
T}})^\frac{5}{2}}(1-e^{2}_{{\rm S}})^{\frac{1}{2}}[(4+3e^{2}_{{\rm
S}})\sin{g_{{\rm T}}}+\nonumber\\
& &+6(x_{{\rm S}}y_{{\rm S}}\cos{g_{{\rm T}}}+y_{{\rm
S}}^{2}\sin{g_{{\rm T}}})]-\nonumber\\
& &
-[\frac{(1-e^{2}_{{\rm S}})^{\frac{1}{2}}}{(1-e^{2}_{{\rm
T}})^{\frac{3}{2}}}+\frac{25}{8}\gamma\frac{3+2e^{2}_{{\rm
T}}}{(1-e^{2}_{{\rm T}})^{3}}(1-\nonumber\\
& &
-\frac{3}{2}e^{2}_{{\rm S}})]y_{{\rm
S}}+\frac{15}{8}\gamma\frac{e^{2}_{{\rm T}}}{(1-e^{2}_{{\rm
T}})^{3}}[y_{{\rm S}}\cos{2g_{{\rm T}}}-\nonumber\\
& &
-x_{{\rm S}}\sin{2g_{{\rm T}}}-\frac{y_{{\rm S}}}{2}(x^{2}_{{\rm
S}}+3y^{2}_{{\rm S}})\cos{2g_{{\rm T}}}+\nonumber\\
& & +x_{{\rm S}}(x^{2}_{{\rm S}}+2y^{2}_{{\rm S}})\sin{2g_{{\rm T}}}]\\
\frac{{\rm d}y_{{\rm S}}}{{\rm d}\tau} & = &
-\frac{5}{16}\alpha\frac{e_{{\rm T}}}{(1-e^{2}_{{\rm
T}})^\frac{5}{2}}(1-e^{2}_{{\rm S}})^{\frac{1}{2}}[(4+3e^{2}_{{\rm
S}})\cos{g_{{\rm T}}}+\nonumber\\
& &+6(x_{{\rm S}}y_{{\rm S}}\sin{g_{{\rm T}}}+x_{{\rm
S}}^{2}\cos{g_{{\rm T}}})]+\nonumber\\
& &
+[\frac{(1-e^{2}_{{\rm S}})^{\frac{1}{2}}}{(1-e^{2}_{{\rm
T}})^{\frac{3}{2}}}+\frac{25}{8}\gamma\frac{3+2e^{2}_{{\rm
T}}}{(1-e^{2}_{{\rm T}})^{3}}(1-\nonumber\\
& &
-\frac{3}{2}e^{2}_{{\rm S}})]x_{{\rm
S}}+\frac{15}{8}\gamma\frac{e^{2}_{{\rm T}}}{(1-e^{2}_{{\rm
T}})^{3}}[x_{{\rm S}}\cos{2g_{{\rm T}}}+\nonumber\\
& &
+y_{{\rm S}}\sin{2g_{{\rm T}}}-\frac{x_{{\rm S}}}{2}(y^{2}_{{\rm
S}}+3x^{2}_{{\rm S}})\cos{2g_{{\rm T}}}-\nonumber\\
& & -y_{{\rm S}}(y^{2}_{{\rm S}}+2x^{2}_{{\rm S}})\sin{2g_{{\rm T}}}]\\
\frac{{\rm d}g_{{\rm T}}}{{\rm d}\tau} & = & \frac{\beta
(2+3e^{2}_{{\rm S}})}{2(1-e^{2}_{{\rm T}})^{2}}-\frac{5}{16}\frac{\alpha \beta
(1+4e^{2}_{{\rm T}})}{e_{{\rm T}}(1-e^{2}_{{\rm
T}})^{3}}(4+3e^{2}_{{\rm S}})\times\nonumber\\
& &
\times(x_{{\rm S}}\cos{g_{{\rm T}}}+y_{{\rm S}}\sin{g_{{\rm T}}})+\frac{5}{8}\beta\gamma\times\nonumber\\
& &
\times\frac{(1-e^{2}_{{\rm S}})^{\frac{1}{2}}}{(1-e^{2}_{{\rm
T}})^{\frac{7}{2}}}[5e^{2}_{{\rm S}}(11+4e^{2}_{{\rm
T}})+3(1+2e^{2}_{{\rm T}})\times\nonumber\\
& & \times((x^{2}_{{\rm S}}-y^{2}_{{\rm S}})\cos{2g_{{\rm
T}}}+2x_{{\rm S}}y_{{\rm S}}\sin{2g_{{\rm T}}})]\label{gtd}\\
\frac{{\rm d}e_{{\rm T}}}{{\rm d}\tau} & = & \frac{5}{16}\frac{\alpha
\beta}{(1-e^{2}_{{\rm T}})^{2}}(4+3e^{2}_{{\rm S}})(y_{{\rm
S}}\cos{g_{{\rm T}}}-\nonumber\\
& &
-x_{{\rm S}}\sin{g_{{\rm T}}})-\frac{15}{8}\beta\gamma\frac{e_{{\rm
T}}(1-e^{2}_{{\rm S}})^{\frac{1}{2}}}{(1-e^{2}_{{\rm T}})^{\frac{5}{2}}}\times\nonumber\\
& & \times(2x_{{\rm S}}y_{{\rm S}}\cos{2g_{{\rm T}}}-(x^{2}_{{\rm
S}}-y^{2}_{{\rm S}})\sin{2g_{{\rm T}}})
\end{eqnarray}
where
\begin{displaymath}
x_{{\rm S}}=e_{{\rm S}}\cos{g_{{\rm S}}},\hspace{0.5cm}y_{{\rm S}}=e_{{\rm
S}}\sin{g_{{\rm S}}},
\end{displaymath}
\begin{displaymath}
\alpha =\frac{m_{1}-m_{2}}{m_{1}+m_{2}}\frac{a_{{\rm S}}}{a_{{\rm T}}},\hspace{0.2cm}\beta
=\frac{m_{1}m_{2}M^{\frac{1}{2}}}{m_{3}(m_{1}+m_{2})^{\frac{3}{2}}}(\frac{a_{{\rm
S}}}{a_{{\rm T}}})^{\frac{1}{2}},
\end{displaymath}
\begin{displaymath}
\gamma=\frac{m_{3}}{M^{\frac{1}{2}}(m_{1}+m_{2})^{\frac{1}{2}}}(\frac{a_{{\rm
S}}}{a_{{\rm T}}})^{\frac{3}{2}}\hspace{1cm}
\mbox{and}
\end{displaymath}
\begin{displaymath}
{\rm d}\tau=\frac{3}{4}\frac{G^{\frac{1}{2}}m_{3}a^{\frac{3}{2}}_{{\rm
S}}}{a^{3}_{{\rm T}}(m_{1}+m_{2})^{\frac{1}{2}}}{\rm d}t.
\end{displaymath} 
After integrating the above averaged equations of motion for
reasonable sets of parameters (e.g. ${m_{1}=0.333}$, ${m_{2}=0.667}$,
${m_{3}=1}$, ${a_{{\rm S}}=1}$ and ${a_{{\rm T}}=10}$), using a 4th-order Runge-Kutta method with  variable
stepsize (Press et al. 1996), it was noticed that ${e_{{\rm T}}}$ remained
almost constant.  If that approximation is taken as an assumption, and
terms 
of order ${e^{2}_{{\rm S}}}$ and ${e^{2}_{{\rm T}}}$ are neglected and only the dominant term 
is retained in equation (\ref{gtd}) (the dominant term is proportional
to ${\beta}$, while the next order term is proportional to
${\alpha\beta}$, which, for the range of parameters discussed in this
paper, is rather small compared to the dominant term), then the system can be reduced to one that can be
solved analytically:
\begin{eqnarray}
\frac{{\rm d}x_{{\rm S}}}{{\rm d}\tau} & = & -By_{{\rm S}}+C\sin{g_{{\rm T}}}\nonumber\\
\frac{{\rm d}y_{{\rm S}}}{{\rm d}\tau} & = & Bx_{{\rm S}}-C\cos{g_{{\rm T}}}
\label{dior}\\
\frac{{\rm d}g_{{\rm T}}}{{\rm d}\tau} & = & A\nonumber,
\end{eqnarray} 
where
\begin{displaymath}
A=\beta,\hspace{0.3cm}
B=1+\frac{75}{8}\gamma,\hspace{0.3cm}
C=\frac{5}{4}\alpha e_{{\rm T}}.
\end{displaymath}
In the limit ${m_{1}>>m_{2}}$ and ${m_{1}>>m_{3}}$, the above system
of equations is in agreement with the corresponding equations of the classical secular planetary theory (Brouwer
${\&}$ Clemence 1961, Murray ${\&}$ Dermott 1999).

The solution to system (\ref{dior}) is:
\begin{eqnarray}
x_{{\rm S}}(\tau) & = &
(K_{1}+\frac{C}{A-B}\cos{{g_{{\rm
T}}}_{0}})\cos{B\tau}+(K_{2}- \nonumber \\
& &
-\frac{C}{A-B}\frac{A}{B}\sin{{g_{{\rm T}}}_{0}})\sin{B\tau}-\frac{C}{A-B}\times\nonumber\\
& & \times\cos{(A\tau+{g_{{\rm T}}}_{0})}\label{secsol1} \\
y_{{\rm S}}(\tau) & = & (K_{1}+\frac{C}{A-B}\cos{{g_{{\rm T}}}_{0}})\sin{B\tau}+\nonumber\\
& & +(\frac{C}{A-B}\frac{A}{B}\sin{{g_{{\rm T}}}_{0}}-K_{2})\cos{B\tau}-\nonumber\\
& & -\frac{C}{A-B}\sin{(A\tau+{g_{{\rm T}}}_{0})},
\label{secsol2}
\end{eqnarray}
where
\begin{math}
K_{1}, K_{2}
\end{math}
are constants of integration and 
\begin{math}
{g_{{\rm T}}}_{0}
\end{math}
is the initial value of ${g_{{\rm T}}}$. 

\subsubsection{Calculation of the initial outer secular eccentricity}

The only thing that remains now is to get an estimate for the initial
${e_{{\rm T}}}$ (since we saw earlier that ${e_{{\rm
T}}}$ remains almost constant, i.e. the outer eccentricity does not
demonstrate any significant long term evolution) and in order to do
that, as was seen in section (\ref{theo}), we need to find an
expression for the short period outer eccentricity.  This can be achieved by following the same procedure as we did in section
\ref{s1}, but this time we do it for the outer orbit.  The equation of
motion of the outer binary is 
\begin{equation}
\ddot{\bmath{R}}=-GM\left(\mu_{1}\frac{\bmath{R}+\mu_{2}\bmath{r}}{|\bmath{R}+\mu_{2}\bmath{r}|^{3}}+\mu_{2}\frac{\bmath{R}-\mu_{1}\bmath{r}}{|\bmath{R}-\mu_{1}\bmath{r}|^{3}}\right)
\end{equation}
and eventually, we obtain, to leading order, for the components of the outer short-period
eccentric vector:
\begin{eqnarray}
x_{2} & = &
\frac{3}{4}\frac{m_{1}m_{2}}{(m_{1}+m_{2})^{\frac{4}{3}}M^{\frac{2}{3}}}\frac{1}{X^{\frac{4}{3}}}\cos{(n_{2}t+\phi)}+\nonumber\\
& & +C_{x_{2}}\label{e21}\\
y_{2} & = &
\frac{3}{4}\frac{m_{1}m_{2}}{(m_{1}+m_{2})^{\frac{4}{3}}M^{\frac{2}{3}}}\frac{1}{X^{\frac{4}{3}}}\sin{(n_{2}t+\phi)}+\nonumber\\
& & +C_{y_{2}}\label{e22}.
\end{eqnarray}
Suppose now that the outer secular eccentric vector is
${\bmath{e}_{{\rm T}}=(x_{{\rm T}},y_{{\rm T}})}$.  Then, the constants
${C_{{\rm x}_{2}}}$ and ${C_{{\rm y}_{2}}}$ in equations (\ref{e21}) and
(\ref{e22}) can be replaced by ${e_{{\rm T}_{1}}}$ and ${e_{{\rm T}_{2}}}$, since
the latter vary slowly compared to ${x_{2}}$ and ${y_{2}}$.
Considering that the outer binary is initially circular, i.e. ${e_{{\rm
out}}=0}$, we obtain:
\begin{eqnarray}
x_{{\rm T}} & = &
-\frac{3}{4}\frac{m_{1}m_{2}}{(m_{1}+m_{2})^{\frac{4}{3}}M^{\frac{2}{3}}}\frac{1}{X^{\frac{4}{3}}}\cos{\phi}\\
y_{{\rm T}} & = &
-\frac{3}{4}\frac{m_{1}m_{2}}{(m_{1}+m_{2})^{\frac{4}{3}}M^{\frac{2}{3}}}\frac{1}{X^{\frac{4}{3}}}\sin{\phi}
\end{eqnarray}
and
\begin{equation}
e_{{\rm T}}=\frac{3}{4}\frac{m_{1}m_{2}}{(m_{1}+m_{2})^{\frac{4}{3}}M^{\frac{2}{3}}}\frac{1}{X^{\frac{4}{3}}}.
\end{equation}
\subsection{A formula for the inner eccentricity}
In paragraphs \ref{s1} and \ref{s2} we derived expressions for the
short period and secular contribution to the inner eccentric vector.
These can be combined to give an expression for the total eccentricity
in the same way we got an estimate for the outer secular eccentricity,
i.e. by replacing the constants in equations (\ref{e11}) and
(\ref{e12}) by equations (\ref{secsol1}) and (\ref{secsol2}), since
the latter evolve on a much larger timescale.  This yields:
\begin{eqnarray}
x_{{\rm in}} & = & x_{1}-C_{{\rm x}_{1}}+x_{{\rm S}}\label{xtot}\\
y_{{\rm in}} & = & y_{1}-C_{{\rm y}_{1}}+y_{{\rm S}}\label{ytot}
\end{eqnarray}
The constants
${K_{1}}$ and ${K_{2}}$ in equations (\ref{secsol1}) and
(\ref{secsol2}) are determined by the fact that the inner eccentricity
is initially zero and are found to be
\begin{eqnarray}
K_{1} & = & \frac{m_{3}}{M}\frac{1}{X^{2}}(\frac{1}{2}-\frac{5}{2}\cos{2\phi}-\frac{15}{16}X^{\frac{1}{3}}m_{*}\times\nonumber\\
& & \times\cos{\phi})\\
K_{2} & = & \frac{m_{3}}{M}\frac{1}{X^{2}}(2\sin{2\phi}+\frac{15}{16}X^{\frac{1}{3}}m_{*}\sin{\phi})+\nonumber\\
& & +\frac{C}{B}\sin{{g_{{\rm T}}}_{0}}.
\end{eqnarray}
We are now able to obtain an expression for the inner eccentricity.
Averaging over time and over the initial relative phase ${\phi}$, the
averaged square inner eccentricity will be given by:
\begin{displaymath}
\overline{e_{{\rm in}}^{2}}=<x^{2}_{{\rm in}}+y^{2}_{{\rm in}}>=\frac{m_{3}^{2}}{M^{2}}\frac{1}{X^{4}}(\frac{43}{4}+\frac{225}{128}m^{2}_{*}X^{\frac{2}{3}})+
\end{displaymath}
\begin{equation}
+\frac{15}{8}\frac{m_{3}}{M}\frac{m_{*}}{X^{\frac{5}{3}}}\frac{C}{A-B}+2\left(\frac{C}{A-B}\right)^{2}.\label{final}
\end{equation}
It should be pointed out here, that the above formula is expected to
be rather inaccurate (in fact, it produces an overestimate for the inner eccentricity) in
situations where the system parameters yield very small values for the
quantity ${A-B}$, i.e. when we are near to a secular resonance, since, as
seen from system (\ref{dior}), ${A}$ and ${B}$ are the secular
frequencies of the inner and outer pericentres respectively.  The
parameters of a resonant system should satisfy the equation ${A-B=0}$,
which yields:
\begin{displaymath}
\frac{m_{1}m_{2}M^{\frac{1}{2}}}{m_{3}(m_{1}+m_{2})^{\frac{3}{2}}}(\frac{a_{{\rm
S}}}{a_{{\rm T}}})^{\frac{1}{2}}-1-\frac{75}{8}\frac{m_{3}}{M^{\frac{1}{2}}(m_{1}+m_{2})^{\frac{1}{2}}}\times
\end{displaymath}
\begin{equation}
\times(\frac{a_{{\rm S}}}{a_{{\rm T}}})^{\frac{3}{2}}=0.  
\end{equation}
None the less, in this case, one could use a formula which only accounts
for the short term evolution of the inner eccentricity, but the
formula will be valid only within a few outer orbital periods.  In this
context, the formula is:
\begin{equation}
\overline{e_{{\rm in}}^{2}}=\frac{m_{3}^{2}}{M^{2}}\frac{1}{X^{4}}(\frac{43}{4}+\frac{225}{128}m^{2}_{*}X^{\frac{2}{3}}).
\end{equation} 

\subsection{Special case: Equal inner binary masses}
In this case, there will be no contribution to the inner eccentricity
from the ${P_{3}}$ term (short period and secular).  The eccentricity
will be dominated by short period terms and the secular contribution
is insignificant compared to that of the short period terms (the only secular contribution to
the inner eccentricity comes from the ${Q_{3}}$ term and is
proportional to ${e^{2}_{{\rm T}}e_{{\rm S}}}$ to leading order).

Following the same procedure as in the more general case
(differentiating the eccentric vector etc.), the components ${x_{1}}$
and ${y_{1}}$ of the eccentric vector are (retaining the two
leading terms):
\begin{equation}
x_{1}=\frac{m_{3}}{M}\frac{1}{X^{2}}(P_{{\rm x21}}(t)+\frac{1}{X}P_{{\rm x22}}(t))+C_{{\rm
x}_{1}}
\label{eq29}
\end{equation}
\begin{equation}
y_{1}=\frac{m_{3}}{M}\frac{1}{X^{2}}(P_{{\rm y21}}(t)+\frac{1}{X}P_{{\rm y22}}(t))+C_{{\rm
y}_{1}}
\label{eq210}
\end{equation}
where
\begin{eqnarray}
P_{{\rm x21}}(t) & = & -\frac{1}{2}\cos{n_{1}t}+\frac{1}{4}\cos{((3n_{1}-2n_{2})t-2\phi)}+\nonumber\\
& & +\frac{9}{4}\cos{((n_{1}-2n_{2})t-2\phi)}\\
P_{{\rm x22}}(t) & = & \frac{1}{6}\cos{((3n_{1}-2n_{2})t-2\phi)}+\nonumber\\
& & +\frac{9}{2}\cos{((n_{1}-2n_{2})t-2\phi)}\\
P_{{\rm y21}}(t)  & = & -\frac{1}{2}\sin{n_{1}t}+\frac{1}{4}\sin{((3n_{1}-2n_{2})t-2\phi)}-\nonumber\\
& & -\frac{9}{4}\sin{((n_{1}-2n_{2})t-2\phi)}\\
P_{{\rm y22}}(t) & = & \frac{1}{6}\sin{((3n_{1}-2n_{2})t-2\phi)}-\nonumber\\
& & -\frac{9}{2}\sin{((n_{1}-2n_{2})t-2\phi)}
\end{eqnarray}
and
\begin{eqnarray}
C_{{\rm x}_{1}} & = &\frac{m_{3}}{M}\frac{1}{X^{2}}
(\frac{1}{2}-\frac{5}{2}\cos{2\phi}-\frac{14}{3}\frac{1}{X}\cos{2\phi})\\
C_{{\rm y}_{1}} & = & \frac{m_{3}}{M}\frac{1}{X^{2}}(-2\sin{2\phi}-\frac{13}{3}\frac{1}{X}\sin{2\phi}).
\end{eqnarray}
The expression for the averaged square eccentricity in this case is:
\begin{equation}
\overline{e_{{\rm in}}^{2}}=\frac{m_{3}^{2}}{M^{2}}\frac{1}{X^{4}}(\frac{43}{4}+\frac{122}{3}\frac{1}{X}).
\label{em}	
\end{equation} 
It is worth mentioning that the term proportional to
${\frac{1}{X^{5}}}$ in equation (\ref{em}) was neglected in the more
general case of the previous section.

\section{COMPARISON WITH OTHER RESULTS}
Eggleton ${\&}$ Kiseleva (1996), in the context of stellar and dynamical
evolution of triple stars, and based on results from numerical
integrations of coplanar, prograde and initially circular orbits,
derived the following empirical formula for the inner mean eccentricity:
\begin{equation}
\bar{e}_{{\rm in}}=\frac{A}{X^{1.5}\sqrt{X-B}},
\label{fegg}
\end{equation}
where ${A}$ and ${B}$ depend on the mass ratios.  For three equal
masses ${A=1.167}$ and ${B=3.814}$.

Equation (\ref{fegg}) can be expanded to first order in terms of ${\frac{1}{X}}$, yielding
\begin{equation}
\bar{e}_{{\rm in}}=\frac{A}{X^{2}}(1+\frac{1}{2}\frac{B}{X}).
\end{equation}  
Using equations (\ref{eq29}) and (\ref{eq210}), for the case of three
equal masses, we get:
\begin{equation}
\bar{e}_{{\rm in}}=\frac{1.157}{X^{2}}(1+\frac{1}{2}\frac{3.816}{X}),
\label{myf}
\end{equation}  
which is in good agreement with the results of Eggleton and Kiseleva.

It is worth mentioning here that, although equation (\ref{fegg}) was
found by Eggleton and Kiseleva to give good results for some mass ratios, it does not for some other.  The
explanation for this could be that, the dominant contribution to the
eccentricity comes from the ${P_{3}}$ term with a factor of
${X^{-\frac{5}{3}}}$ and not from the ${P_{2}}$ term, as one might
expect (see section \ref{s1}).

\section{COMPARISON WITH NUMERICAL RESULTS}
In order to test the validity of the formulae derived in the previous
sections, we integrated the full equations of motion numerically, using
a symplectic integrator with time transformation (Mikkola 1997).

The code calculates the relative position and velocity vectors of the
two binaries at every time step.  Then, by using standard two body formulae,
we computed the orbital elements of the two binaries.
The various parameters used by the code, were given the following
values: writing index ${Iwr=1}$, average number of steps per inner binary
period ${NS=60}$, method coefficients ${a1=1}$ and ${a2=15}$,
correction index ${icor=1}$.
In all simulations, we confined ourselves to systems with mass ratios within the
range ${10:1}$ since, among stellar triples, mass ratios are rare
outside a range of approximately ${10:1}$, although such systems would be
inherently difficult to recognise (Eggleton ${\&}$ Kiseleva 1995); and
initial period ratio ${X \geq 10}$.  We also used units such that
${G=1}$ and ${m_{1}+m_{2}=1}$ and we always started the integrations
with ${a_{1}=1}$.  In that system of units, the initial conditions for the numerical
integrations were as follows:
\begin{displaymath}
r_{1}=1,\hspace{0.5cm} r_{2}=0,\hspace{0.5cm} r_{3}=0
\end{displaymath}	
\begin{displaymath}
R_{1}=a_{2}\cos{\phi},\hspace{0.5cm} R_{2}=a_{2}\sin{\phi},\hspace{0.5cm} R_{3}=0
\end{displaymath}	
\begin{displaymath}
\dot{r}_{1}=0,\hspace{0.5cm} \dot{r}_{2}=1,\hspace{0.5cm} \dot{r}_{3}=0
\end{displaymath}	
\begin{displaymath}
\dot{R}_{1}=-\sqrt{\frac{M}{a_{2}}}\sin{\phi},\hspace{0.5cm} \dot{R}_{2}=\sqrt{\frac{M}{a_{2}}}\cos{\phi},\hspace{0.5cm} \dot{R}_{3}=0,
\end{displaymath}	
where ${\bmath {r}}$ and ${\bmath {R}}$ are the relative
position vectors of the inner and outer orbit respectively.

\subsection{SHORT PERIOD EFFECTS} 

First we tested the validity of equations (\ref{e11}) and
(\ref{e12}).  The integrations and comparison with the analytical results were done for
${\phi=90^{\circ}}$, i.e. the outer binary was ahead of the inner one
at right angles.  However, this does not affect the qualitative
understanding of the problem at all. 
 
\begin{figure}
\begin{center}
\includegraphics[width=80mm,height=60mm]{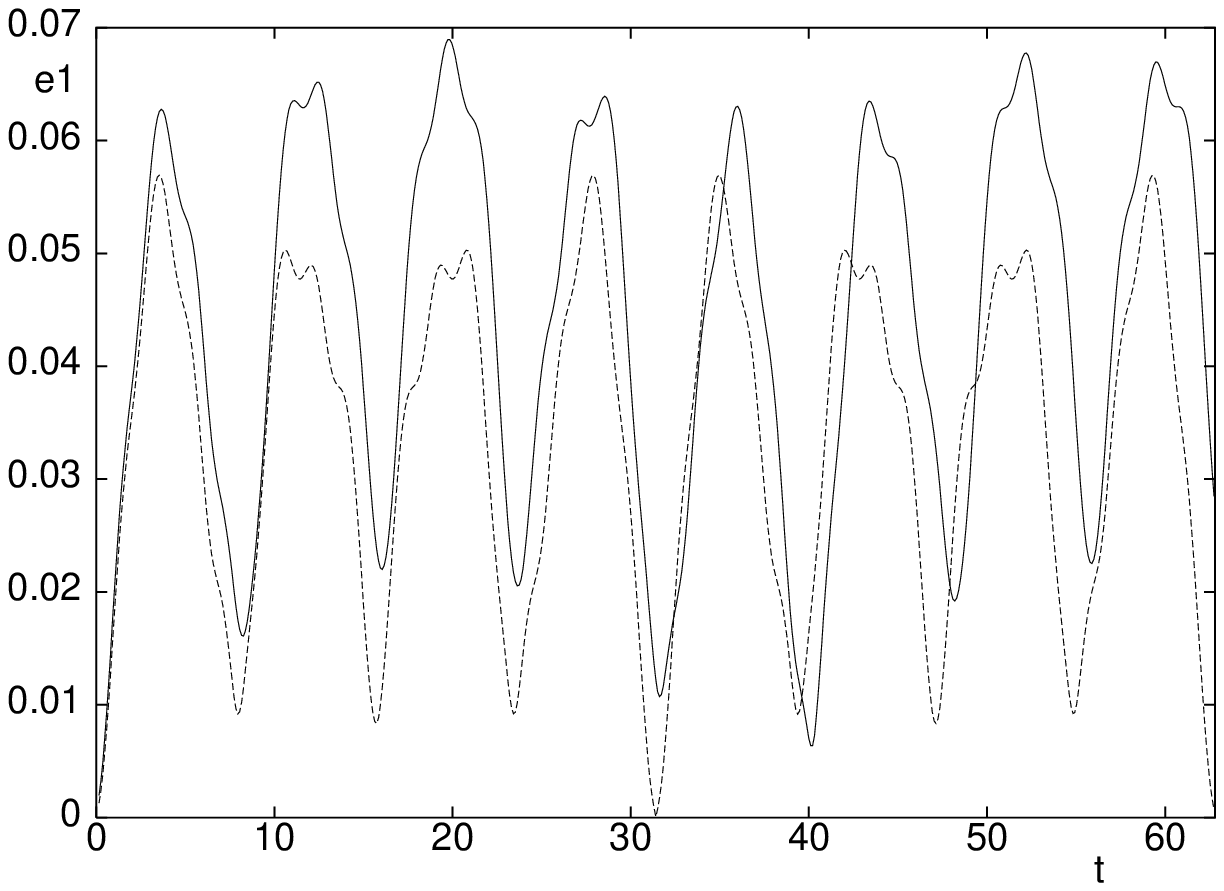}
\caption[]{Inner eccentricity against time for a system with
${m_{1}=0.5}$, ${m_{3}=5}$, ${X_{0}=10}$ and ${\phi=90^{\circ}}$.  The integration 
time span is one outer orbital period (${T_{{\rm out}}=62.8}$).  The
continuous curve comes from the numerical integration of the full equations of motion, while the
dashed curve is a plot of equations (\ref{e11}) and (\ref{e12}).  In
the system of units used, the inner binary period is ${T_{{\rm in}}=2\pi}$.}
\label{fig2}
\end{center}
\end{figure}

\begin{figure}
\begin{center}
\includegraphics[width=80mm,height=60mm]{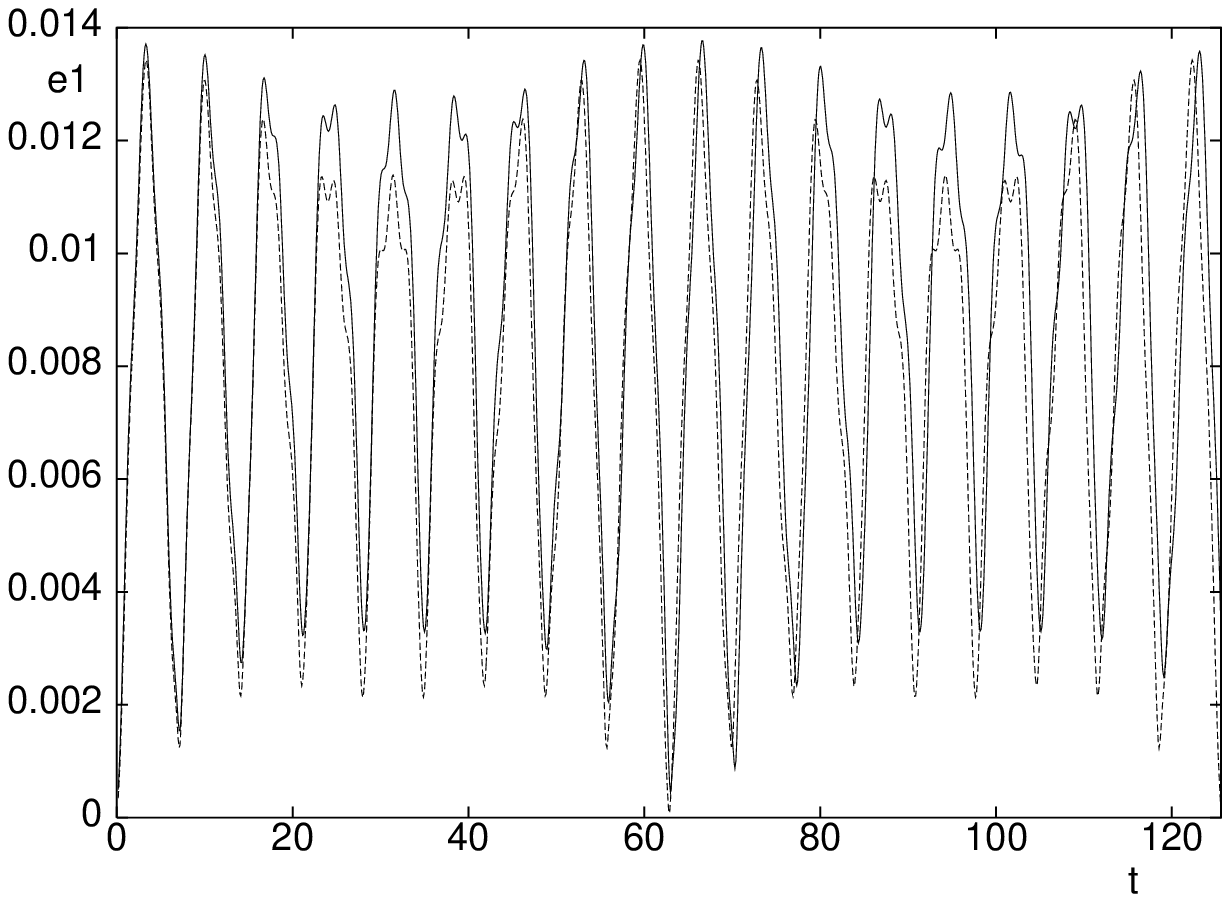}
\caption[]{Inner eccentricity against time for a system with
${m_{1}=0.5}$, ${m_{3}=5}$, ${X_{0}=20}$ and ${\phi=90^{\circ}}$.  The
integration time span is one outer orbital period (${T_{{\rm out}}=125.6}$).  The continuous curve comes from 
the numerical integration of the full equations of motion, while the
dashed curve is a plot of equations (\ref{e11}) and (\ref{e12}).  In
the system of units used, the inner binary period is ${T_{{\rm in}}=2\pi}$.}
\label{fig25}
\end{center}
\end{figure}

\begin{figure}
\begin{center}
\includegraphics[width=80mm,height=60mm]{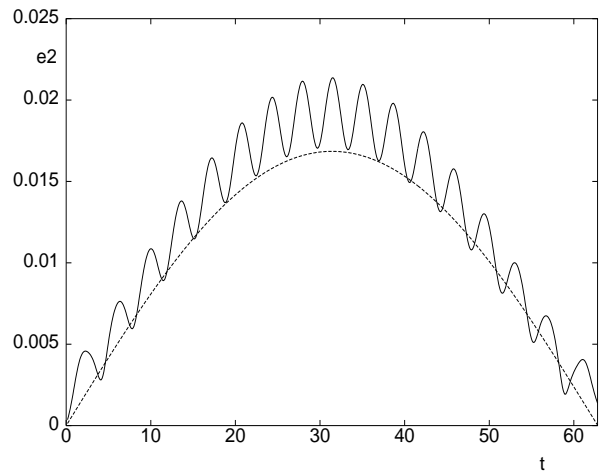}
\caption[]{Outer eccentricity against time for a system with
${m_{1}=0.5}$, ${m_{3}=0.05}$, ${X_{0}=10}$ and ${\phi=90^{\circ}}$ .  The integration 
time span is one outer orbital period (${T_{{\rm out}}=62.8}$).  The
continuous curve comes from the numerical integration of the full equations of motion, while the
dashed curve is a plot of equations (\ref{e21}) and (\ref{e22}).  In the system of units used, the inner binary period is ${T_{{\rm in}}=2\pi}$.}
\label{fig3}
\end{center}
\end{figure}

The results are presented in Table 1, which gives the percentage error between the
averaged, over time, numerical and theoretical ${e_{{\rm in}}}$ (the theoretical eccentricity was obtained by
evaluating  equations (\ref{e11}) and (\ref{e12}) everytime we
had an output from the symplectic integrator; both averaged
numerical and theoretical eccentricities  were calculated by using the
trapezium rule).  The
integrations were performed over one outer orbital period time span (in our
system of units, the initial outer orbital period is ${T_{{\rm out}}=2 \pi
X_{0}}$, where ${X_{0}}$ is the initial period ratio).
For each pair ${(m_{3},X_{0})}$ in Table 1, there are five entries,
corresponding, from top to bottom, to the following inner binaries:
${m_{1}=0.1-m_{2}=0.9}$, ${m_{1}=0.2-m_{2}=0.8}$,
${m_{1}=0.3-m_{2}=0.7}$, ${m_{1}=0.4-m_{2}=0.6}$ and
${m_{1}=0.5-m_{2}=0.5}$.  A dash in Table 1 denotes that the analogy among the masses was
outside the range ${10:1}$.  The results show a rather significant
error for systems with strong perturbation to the inner binary (small
${X_{0}}$-large ${m_{3}}$).  However, the error drops considerably as we move to
larger values of ${X_{0}}$ (the error becomes less or close to ${10 \%}$
for ${X_{0}=20}$).  This is consistent with our aim to obtain
a reasonable model for the evolution of the inner eccentricity in
hierarchical triple systems with well separated components.  One
should bear in mind that a period ratio of ${20}$ is considered, as
seen in the introduction, to be close to the
lower boundary for a hierarchical triple system.  Fig. \ref{fig2} is a plot of inner binary eccentricity against time for a system with
${m_{1}=0.5}$, ${m_{3}=5}$, ${X_{0}=10}$ and ${\phi=90^{\circ}}$.  The
continuous curve has been produced as a result of the numerical integration of
the full equations of motion, while the dashed curve is based on
equations (\ref{e11}) and (\ref{e12}).  It is quite obvious that the
theory does not work very well for that parameter combination (also
see Table 1).  One should note that the maximum eccentricity is of ${O(10^{-2})}$,
implying that the inner orbit is close to a circle.  This represents
an extreme case of the triple systems studied in this paper, in the
sense that the perturbation to the inner binary is strong and
therefore, the rest of the systems investigated in this paper, would
be expected to have a maximum inner eccentricity of the same or
smaller order.  Fig. \ref{fig25} demonstrates the
inner eccentricity  evolution of the same system as Fig. \ref{fig2},
but for ${X_{0}=20}$.  The improvement in the theory, as ${X_{0}}$
increased, is demonstrated by the good agreement between the numerical
(continuous curve) and the theoretical result (dashed curve).
Finally, Fig. \ref{fig3} shows a similar situation as Fig. \ref{fig2}
(i.e. strong perturbation), but for
the outer binary.  The parameters of the system are the same as
in Fig. \ref{fig2} except ${m_{3}=0.05}$ this time.  Note the
satisfactory agreement between the numerical (continuous curve) and
the theoretical result (dashed curve based on equations (\ref{e21}) and
(\ref{e22})), although our intention was to compute only the dominant contribution to
the outer short period eccentricity.  Again, the outer
orbit could be approximated by a circle.  Hence, the assumption of
circular orbits in section (\ref{s1}) is well justified.  That was also
confirmed when equation (\ref{roo5}) was numerically tested against
the rate of change of equation (\ref{ecve}).  For instance for a
system with ${m_{1}=0.4}$, ${m_{3}=4}$, ${X_{0}=10}$ and
${\phi=90^{\circ}}$, for which the perturbation to the inner binary is
rather strong, the absolute percentage error between the magnitudes of
the exact and the approximate rate of change of the inner eccentric
vector oscillated between ${0 - 5\%}$ with a period of approximately half an outer orbital period (the integration time span was
${1000}$ outer orbital period).  That oscillation interval was reduced to
${0 - 1\%}$ when ${X_{0}=20}$ and the same reduction analogy (about ${80\%}$)
was also observed in the maximum eccentricity of the two systems,
which was consistent with the fact that the error in equation (\ref{roo5}) was of order
${O(e)}$.  Hence, the analytical predictions for
the eccentricity made by equations (\ref{e11}) and (\ref{e12}) should
not break down for long evolution timescales. 

\begin{table}
\caption[]{Percentage error between the
averaged numerical and averaged theoretical ${e_{{\rm in}}}$.  The
theoretical model is based on equations (\ref{e11}) and
(\ref{e12}).  For all systems, ${\phi=90^{\circ}}$.}
\vspace{0.1 cm}
\begin{center}	
{\footnotesize \begin{tabular}{c c c c c c c c c c}\hline
${m_{3}\backslash\ X_{0}}$ & ${10}$ & ${15}$ & ${20}$ & ${25}$ & ${30}$ & ${50}$ \\
\hline
0.05 & - & - & - & - & - & - \\
     & - & - & - & - & - & - \\
     & - & - & - & - & - & - \\
     & - & - & - & - & - & - \\
     & 6.2 & 2.9 & 1.7 & 1.2 & 0.9 & 0.4 \\
0.09 & 18.6 & 11.7 & 8.5 & 6.6 & 5.4 & 3 \\
     & 19.3 & 12.3 & 8.9 & 6.9 & 5.7 & 3.2 \\
     & 19.8 & 12.7 & 9.3 & 7.3 & 6 & 3.4 \\
     & 20.1 & 13 & 9.5 & 7.5 & 6.2 & 3.6 \\
     & 6.8  & 3.4 & 2.1 & 1.4 & 1.1 & 0.5 \\
0.5  & 24.1 & 15.5 & 11.3 & 8.9 & 7.3 & 4.3 \\
     & 24.5 & 15.7 & 11.5 & 9.1 & 7.4 & 4.3 \\
     & 24.6 & 15.8 & 11.6 & 9.2 & 7.5 & 4.4 \\
     & 24.6 & 15.9 & 11.7 & 9.2 & 7.6 & 4.4 \\
     & 12   & 6.5  & 4.3  & 3.2 & 2.5 & 1.3 \\
 1   & 27.9 & 18   & 13.2 & 10.4 & 8.6 & 5 \\
     & 28   & 18   & 13.2 & 10.4 & 8.6 & 5  \\
     & 28   & 18   & 13.1 & 10.4 & 8.6 & 5  \\
     & 27.8 & 17.9 & 13.1 & 10.3 & 8.5 & 5  \\
     & 15.8 & 8.8  & 5.9  & 4.4  & 3.4 & 1.8 \\
1.5  & - & - & - & - & - & - \\
     & 30.1 & 19.4 & 14.2 & 11.2 & 9.2 & 5.4 \\
     & 30   & 19.3 & 14.1 & 11.1 & 9.2 & 5.4 \\
     & 29.8 & 19.1 & 14   & 11   & 9.1 & 5.3 \\
     & 18.2 & 10.1 & 6.9  & 5.1  & 4   & 2.2 \\
 2   & - & - & - & - & - & - \\
     & 31.6 & 20.3 & 14.9 & 11.7 & 9.6 & 5.7 \\
     & 31.3 & 20.1 & 14.7 & 11.6 & 9.5 & 5.6 \\
     & 31.1 & 20   & 14.6 & 11.5 & 9.4 & 5.5 \\
     & 19.8 & 11.1 & 7.5 & 5.6 & 4.4 & 2.4 \\
2.6  & - & - & - & - & - & - \\
     & - & - & - & - & - & - \\
     & 32.4 & 20.8 & 15.2 & 12   & 9.9 & 5.8 \\
     & 32.4 & 20.7 & 15.1 & 11.9 & 9.8 & 5.7 \\
     & 21.3 & 11.9 & 8    & 6    & 4.7 & 2.5 \\
 3   & - & - & - & - & - & - \\
     & - & - & - & - & - & - \\
     & 33 & 21.2 & 15.5 & 12.1 & 10  & 5.9 \\
     & 33 & 21.1 & 15.4 & 12.1 & 9.9 & 5.8 \\
     & 22 & 12.3 & 8.3  & 6.2  & 5   & 2.6 \\
3.4  & - & - & - & - & - & - \\
     & - & - & - & - & - & - \\
     & - & - & - & - & - & - \\
     & 33.5 & 21.4 & 15.6 & 12.2 & 10 & 5.8 \\
     & 22.6 & 12.6 & 8.5  & 6.4  & 5  & 2.7 \\
 4   & - & - & -  & - & - & - \\
     & - & - & - & - & - & - \\
     & - & - & - & - & - & - \\
     & 34   & 21.7 & 15.8 & 12.4 & 10.2 & 5.9 \\
     & 23.3 & 13   & 8.8  & 6.6 & 5.2 & 2.8 \\
4.5  & - & - & - & - & - & - \\
     & - & - & - & - & - & - \\
     & - & - & - & - & - & - \\
     & - & - & - & - & - & - \\
     & 23.8 & 13.3 & 9 & 6.7 & 5.3 & 2.9 \\
 5   & - & - & - & - & - & - \\
     & - & - & - & - & - & - \\
     & - & - & - & - & - & - \\
     & - & - & - & - & - & - \\
     & 24 & 13.5 & 9.1 & 6.8 & 5.4 & 2.9 \\
\hline
\end{tabular}}
\end{center}	
\end{table}

\subsection{SHORT AND LONG PERIOD EFFECTS} 
Next, we tested equation (\ref{final}), which accounts for the short
period and secular effects to the inner eccentricity.  The formula was
compared with results obtained from integrating the full
equations of motion numerically.  These results are presented in Table 2, which
gives the absolute percentage error between the averaged, over time and initial phase ${\phi}$, numerical
${e^{2}_{{\rm in}}}$ and equation (\ref{final}).  In the case of a system with noticeable secular evolution, the
error is accompanied by the period of the oscillation
of the eccentricity, which is the same as the integration time span,
while the rest of the systems were
integrated over one outer orbital period, since there was not any
noticeable secular evolution when those systems were integrated over
longer time spans.  Each system was numerically integrated for
${\phi=0^{\circ}-360^{\circ}}$ with a step of ${10^{\circ}}$.  After
each run, ${e^{2}_{{\rm in}}}$ was averaged over time using the trapezium
rule and after the integrations for all ${\phi}$ were done, we
averaged over ${\phi}$ by using the rectangle rule.  The integrations
were also done for smaller steps in ${\phi}$ (${1^{\circ}}$ and
${0.1^{\circ}}$), but there was not any difference in the outcome.
All the integrations presented in Table 2 were done for ${m_{1}=0.2}$,
but similar results are expected for the other inner
binary mass ratios.  

\begin{figure}
\begin{center}
\includegraphics[width=80mm,height=60mm]{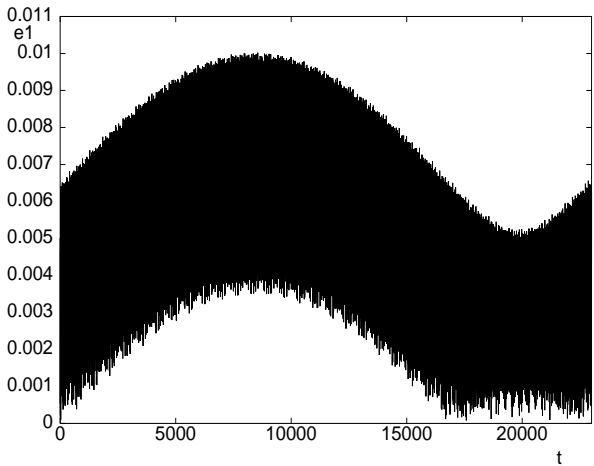}
\caption[]{Secular resonance for a system with ${m_{1}=0.2}$,
${m_{3}=0.09}$, ${X_{0}=10}$ and ${\phi=90^{\circ}}$.  The graph comes from numerical
integration of the full equations of motion.  The two binary periods
are ${T_{{\rm in}}=2\pi}$ and ${T_{{\rm out}}=62.8}$.}
\label{fig4}
\end{center}
\end{figure}

\begin{figure}
\begin{center}
\includegraphics[width=80mm,height=60mm]{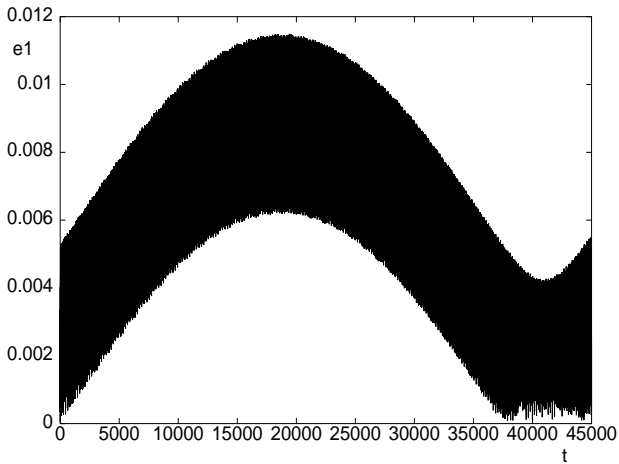}
\caption[]{Inner eccentricity against time for a system with ${m_{1}=0.2}$,
${m_{3}=0.09}$, ${X_{0}=10}$ and ${\phi=90^{\circ}}$  based on
equations (\ref{xtot}) and (\ref{ytot}).  Note the long period and
large amplitude of the oscillation.  The two binary periods
are ${T_{{\rm in}}=2\pi}$ and ${T_{{\rm out}}=62.8}$.}
\label{fig5}
\end{center}
\end{figure}

For a system with ${m_{3}=0.09}$ and ${X_{0}=10}$,
we have a rather large error of ${72.5}$ per cent.  In this case, besides the error that
arises from
the short period terms, there is a significant discrepancy between the
theoretical secular solution and the numerical results, as seen in
Figs. \ref{fig4} and \ref{fig5}.  It is easily noted in Fig. \ref{fig5},
which is a plot based on equations (\ref{xtot}) and (\ref{ytot}), that
the secular period and amplitude of the oscillation are larger than the
ones obtained from the numerical integrations (Fig. \ref{fig4}).  This is due to the fact that the
system is in the vicinity of a secular resonance.  The effect of the resonance gets
less significant as ${X_{0}}$ increases.

\begin{table}
\caption[]{Absolute percentage error between the averaged numerical
${e^{2}_{{\rm in}}}$ and equation (\ref{final}).}
\vspace{0.1 cm}
\begin{center}	
{\footnotesize \begin{tabular}{c c c c c c c c c c}\hline
${m_{3}\backslash\ X_{0}}$ & ${10}$ & ${15}$ & ${20}$ & ${25}$ & ${30}$ & ${50}$ \\
\hline
0.09 &72.5&16.5&3&0.5&2.7&1.9 \\
     &23000&57000&97000&145000&196000&490000\\
0.5  &37.1&25&18.7&14.6&12.2&6.9\\
     & &6000&10000&17000&23000&70000\\
 1   &41.5&27.7&20.5&16.9&13.8&8.2 \\
     & & & &7500&15000&40000\\
1.5  &43.9&29.5&22&17.4&14.3&8.1\\
     & & & & & & \\
 2   &45.4&30.5&22.9&18.2&15.1&8.7\\
     & & & & & & \\
\hline
\end{tabular}}
\end{center}	
\end{table}

\section{CONCLUSION}

We have constructed a method to get an estimate of the inner
eccentricity in hierarchical triple systems on initially circular and
coplanar orbits.  The equations developed throughout this paper, seem
to give reasonable results for the parameter ranges discussed.  This
can be quite important for systems with close inner binaries.  Of
course, it is always possible to improve the theory by adding more
short period terms in the expansion of the eccentric vector.  Our
future aim is to expand that kind of calculation to systems with a
wider range of orbital characteristics, such as systems with
inclined orbits.

\section*{ACKNOWLEDGMENTS}
The author is grateful to Prof. Douglas Heggie for all the useful
discussions on the context of this paper. I also want to thank Seppo
Mikkola, who kindly provided the code for integrating hierarchical
triple systems.
 
\section*{REFERENCES}
Brouwer D., Clemence G. M., 1961, Methods of Celestial
Mechanics. Academic Press, NY\\ 
Eggleton P. P., Kiseleva L. G., 1995, ApJ, 455, 640\\
Eggleton P. P., Kiseleva L. G., 1996, in Wijers R. A. M. J., Davies
M. B., eds, Proc. NATO Adv. Study Inst., Evolutionary Processes in
Binary Stars.  Kluwer Dordrecht, p. 345\\
Fekel F. C., Jr.; Tomkin J., 1982, ApJ 263, 289\\
Ford E. B., Kozinsky B., Rasio F. A., 2000, ApJ, 535, 385\\
Hinkle K. H., Fekel F. C., Johnson D. S., Scharlach W. W. G., 1993, AJ, 105, 1074\\
Jha S., Torres G., Stefanik R. P., Latham D. W., Mazeh T., 2000, MNRAS, 317, 375\\
Kiseleva L. G., Eggleton P. P., Mikkola S., 1998, MNRAS, 300, 292\\
Krymolowski Y., Mazeh T., 1999, MNRAS, 304, 720\\
Marchal C., 1990, The Three-Body Problem.  Elsevier Science Publishers, the Netherlands\\
Mikkola S., 1997, CeMDA, 67, 145\\
Murray C. D., Dermott S. F., 1999, Solar System Dynamics. Cambridge
Univ. Press, Cambridge\\
Press W. H., Teukolsky S. A., Vetterling W. T.,
Flannery B. P., 1996, Numerical Recipes In Fortran 77 (2nd
ed.).  Cambridge Univ. Press, NY
\end{document}